\documentstyle[aps,preprint]{revtex}              
\draft
\title{Low frequency admittance of quantized Hall conductors}
\author{T. Christen and M. B\"uttiker}
\address{Universit\'e de Gen\`eve, 24 Quai E. Ansermet \\ 
CH-1211 Gen\`eve, Suisse}
\begin{document}
\maketitle
\begin{abstract}
We present a current and charge conserving theory 
for the low frequency admittance of a two-dimensional
electron gas connected to ideal metallic contacts and subject to a
quantizing magnetic field. In the framework of the edge-channel picture,
we calculate the admittance up to first order
with respect to frequency. The transport coefficients in first order
with respect to frequency, which are called \em emittances\rm ,
determine the charge emitted into a contact of the sample or a gate 
in response to an oscillating voltage applied to a contact of the sample
or a nearby gate. The emittances depend on the potential distribution
inside the sample which is established in response to the oscillation
of the potential at a contact. 
We show that the emittances can be related to the elements of an
electro-chemical capacitance matrix which describes a (fictitious)
geometry in which each edge channel is coupled to its own reservoir. 
The particular relation of the emittance matrix to this electro-chemical
capacitance matrix depends strongly on the topology of the 
edge channels: We show that edge channels which connect
different reservoirs contribute with a negative capacitance to the
emittance. For example, while the emittance of a two-terminal
Corbino disc is a capacitance, the emittance of a two-terminal quantum
Hall bar is a negative capacitance. The geometry of the edge-channel
arrangement in a many-terminal setup is reflected by symmetry properties
of the emittance matrix. We investigate the effect of voltage probes and
calculate the longitudinal and the Hall
resistances of an ideal four-terminal Hall bar for low frequencies. 
\end{abstract}
\newpage
\section{Introduction}
\label{Intro}
The quantized Hall effect\cite{Klit1} provides particularly interesting tests
of our understanding of electrical transport. 
Application of a resistance formula which treats all contacts to 
a two-dimensional electron gas on equal footing\cite{Buet1} has considerably
revised the traditional picture of the quantized Hall effect and 
has led to the successful explanation of many novel experiments\cite{Haug}.
It is the purpose of this work to approach the 
low frequency electrical transport
in two-dimensional electron systems (2DES) 
subject to strong magnetic fields from a similar point of view. 
In contrast to the dc-transport properties, which have become
increasingly well understood, the ac-transport properties have found much
less attention. However,
a charge and current conserving theory for the low
frequency admittance $G_{\alpha \beta} (\omega) $
of a general arrangement of mesoscopic
conductors has recently been worked out
\cite{Buet7,Buet5}. We apply this theory to
Hall systems in the integer quantum Hall regime at a plateau.
A charge and current conserving theory requires knowledge
of the non-equilibrium potential distribution inside the conductor. 
In the quantum Hall regime the determination of this potential 
becomes simple due to the formation of edge channels \cite{Halp1}.
As discussed in detail by Chklovskii et al.\cite{Chkl1}
and closely related works 
\cite{Coop1,Lier1,Efro1,Chan1,Been1,Gelf1,Mend1}, 
there occurs a decomposition of a 2DES in metal-like
edge channels and dielectric-like regions. Consequently,
the non-equilibrium potential is also
determined by the properties of the edge channels. 
If the edge channels behave like perfect metals they screen any excess
charge. The resulting non-equilibrium potential is determined by the
geometry of the edge-channel arrangement alone. 
On the other hand, if the charge in the edge 
channels is not perfectly screened the non-equilibrium
potential depends on the density of states of the edge 
channels. The resulting potential distribution is not of geometrical
nature alone but contains quantum corrections due to the finite
density of states of the edge channels.
It is of particular interest to investigate to what extent 
such quantum corrections affect the dynamic transport properties 
of a 2DES.\\ \indent
The admittance $G_{\alpha
\beta} (\omega) $ gives the linear current response
$\delta I_{\alpha} \exp(-i\omega t)$ at a contact $\alpha $ of a
device, if at contact $\beta $ a
voltage oscillation $\delta V_{\beta} \exp(-i\omega t)$ is applied:
\begin{equation}
\delta I_{\alpha}(\omega) = \sum _{\beta}
G_{\alpha \beta} (\omega) \; \delta V_{\beta}(\omega)\;\;.
\label{eq1}   
\end{equation} 
The voltage variation $\delta V_{\beta}$ is related to the variation
of the electro-chemical
potential $\delta \mu _{\beta} $ in reservoir $\beta $ by  $\delta
\mu _{\beta}=e\delta V _{\beta}$, where $e$ is the electron charge. 
The theory \cite{Buet7,Buet5} deals with the 
dc-conductance, $G_{\alpha \beta }^{(0)}$,
and the first-order term with respect to frequency,
$ E_{\alpha \beta}\equiv i(dG_{\alpha \beta}/d\omega)_{\omega = 0}$,
which is called the \em emittance \rm matrix. The low frequency 
admittance can then approximately be written in the form  
\begin{equation}
G_{\alpha \beta} (\omega)= G_{\alpha \beta} ^{(0)}-i\omega E_{\alpha \beta}
\;\;.
\label{eq2} 
\end{equation}
For an array of macroscopic conductors of which each is connected
to a single contact, the emittance is just a geometrical capacitance,
i.e. $E_{\alpha \beta }= C_{\alpha \beta }$. However,
this is not true for {\em mesoscopic} conductors and conductors which
connect {\em different} reservoirs \cite{Buet5}. Firstly, it is not
the geometrical capacitance but rather the {\em electro-chemical
capacitance } which relates charges at mesoscopic
conductors with voltage variations
in the reservoirs. Secondly, conductors which connect
different reservoirs allow a transmission of charge which leads to
inductance-like contributions to the emittance.\\ \indent
We shall show
that the emittance $E_{\alpha \beta}$ of a quantized Hall sample
is the sum over 
elements of the electro-chemical capacitance matrix, $c_{\mu, kl}$,
for edge-channels $l$ into which charge is injected at contact
$\beta $ and for edge-channels $k$ from
which charge is emitted into contact $\alpha $. 
The electro-chemical capacitance matrix, $c_{\mu, kl}$, is 
determined by considering each edge channel as a metal strip connected 
to a single contact. Our expression for the emittance is simple
enough in order to discuss arbitrarily complicated edge-channel
arrangements without much technical effort, once the
electro-chemical capacitance matrix of the edge-channel arrangement is known.
We emphasize here that our theory satisfies charge and current conservation
which are due to a perfect screening of
electric fields in the reservoirs and in the gates used to form the conductor. 
Current conservation implies that
the admittance satisfies $\sum _{\beta} G_{\alpha \beta}
= \sum _{\alpha} G_{\alpha \beta} =0$.\\ \indent
Two simple geometries can be used to illustrate the different behavior of
the emittance, namely the Hall bar geometry (Fig. \ref{fig1}.a),
and the Corbino geometry (Fig. \ref{fig1}.b). 
We will show that 
in a Hall bar kinetic charge motion of electrons along 
the edge channels dominates the Coulomb interaction between the
reservoirs. The emittance is a negative electro-chemical
capacitance, i.e. $E= -C_{\mu} ,$ with $C_{\mu} > 0$. On the other hand,
in the Corbino geometry contacts are located at the inner and the outer
perimeter of an annular film\cite{vanW1,Jean1}. Hence, edge channels do
not connect different reservoirs and will thus not contribute
to a dc-current. Moreover, in contrast to the bar geometry
in the Corbino disc capacitive effects
dominate and the emittance is a capacitance, i.e.  $E= C_{\mu}$.\\ \indent 
The transverse potential profile in a cross section of
these conductors is qualitatively shown shown in Fig. \ref{fig2}
which is to be discussed below. Here we only mention that
the similarity of this potential for the two different setups
applies only to the bulk of the sample.
We will assume in this work that the capacitances and emittances are 
dominated by the bulk and that contact capacitances can be neglected. 
\\ \indent

The paper is organized as follows. In Sect. \ref{Equilibrium}
we briefly recall the edge-channel picture of a 2DES at equilibrium.
In Sect. \ref{Nonequilibrium}, we discuss the
dc-nonequilibrium electric potential in terms of
an electro-chemical capacitance matrix, and the 
expression for the dc-conductance $G_{\alpha \beta }^{(0)}$
is derived. In Sect. \ref{Emittance} we outline
the theory of emittances and derive an expression for the emittance matrix
$ E_{\alpha \beta}$ for quantized Hall samples. The result is
applied to various specific examples in Sect.
\ref{Examples}. In Sect. \ref{incoherent} we investigate the effect of a
voltage probe. As an application, we calculate in Sect. \ref{Hall}
the longitudinal resistance and the Hall resistance of a four-probe
quantum Hall bar for low frequencies.   
\section{Quantized Hall Samples at Equilibrium}
\label{Equilibrium}
We begin with a brief discussion of important equilibrium
properties of a 2DES at an (integer) Hall plateau. Consider the two-terminal
quantum Hall bar in Fig. \ref{fig1}.a. The bar
is connected on either side via ideal contacts to particle reservoirs
$\alpha =1,2$ at electro-chemical potentials $\mu_{\alpha} =
E_{F,\alpha}+eU_{\alpha}$. Here, $E_{F,\alpha}$ and $U_{\alpha}$
denote the chemical and the electric potential of reservoir
$\alpha$, respectively. The strong magnetic field is assumed
to be perpendicular to the plane of the 2DES.
Translational invariance of the potential $eU(x)$ in the $y$-direction allows
one to restrict the considerations to a transverse cross section of
the sample. The single-particle potential as a function of
$x$ is sketched in Fig. \ref{fig2}.a for the equilibrium case
where the reservoirs are kept at equal electro-chemical potential,
say $\mu = E_{F}$ where $U_{\alpha}\equiv 0$. For the moment, we assume
that the Fermi level lies in the region
between the extended bulk states of the first and the second (spin-split)
Landau levels. Hence, in the bulk the states of a
single Landau level are
completely filled (black dots in Fig. \ref{fig2}.a).
At the sample boundary, however,
the confinement potential
strongly bends up the single-particle potential which, therefore,
intersects the Fermi energy. This leads to the existence of
extended states at the Fermi level (edge channels) along the
sample boundary.
For non-interacting electrons\cite{Halp1} 
the intersection of the single particle energy with the Fermi energy
is sharp. The transverse size of an edge channel is of the order of 
a magnetic length,
$l_{m}= \sqrt{\hbar /|eB|}$.  
The mean drift velocity of a carrier with coordinate $x$ points in
$y$-direction and is given by \cite{McDo1} $ v (x)= (dU/dx)/B.$
This is just the Lorentz drift of 
the center of a cyclotron orbit in an electric field.
\\ \indent   
In a quantized Hall sample a current density
exists which is a pure equilibrium phenomenon and cannot
lead to a current between reservoirs. For a filled Landau level,
the diamagnetic current density can be written in the form 
\cite{McDo1}
\begin{equation}
j = -\frac{e^{2}}{h}\: \frac{dU}{dx} \;\;.
\label{eq3}
\end{equation}
The total current through a contact is obtained by a transverse
spatial integral of $j(x)$. It vanishes at equilibrium since at both
boundaries of integration,
$\mu = E_{F}+eU_{k}$ holds, where $k=1,2$ labels the edge channels.
Of course, this statement is valid independent
of the geometrical arrangement of the edge channels as long as the
cross section is constructed such that all edge channels of a contact
are included. In particular, it is independent of the  specific
space dependence of the equilibrium potential
which can be very complicated.\\ \indent
The inclusion of Coulomb
interaction, even within a mean-field approximation,
drastically affects the results of the single-particle approach.
Coulomb interactions lead to an electro-static restructuring of the edge
\cite{Chkl1,Coop1,Lier1,Efro1,Chan1,Been1}.
The 2DES is composed of alternating strips of compressible and
incompressible electron liquids with finite widths. Incompressible
regions where the filling factor has discrete values behave like
dielectrics. Quantitative analytical 
predictions of the widths of the compressible and incompressible
strips have been made by Chklovskii et al.\cite{Chkl1}.
These predictions are in good agreement with numerical work by
Lier and Gerhardts\cite{Lier1}.
Edge channels correspond to the compressible regions
where single-particle states are partially filled and 
the electric potential is pinned to the Fermi level (flat parts in 
Fig. \ref{fig2}.a).
Edge channels have screening properties similar to metallic strips. 
The many-particle effects become important if the
strength $ dU/dx$ of the (unscreened)
confinement field is weaker than the characteristic electric
field of electron-electron interaction, i.e.
if $\alpha \equiv |dU/dx| 4\pi \epsilon _{0}\epsilon_{r} l_{m}^{2}/e \ll
1$, where $\epsilon _{0} \epsilon_{r}$ is the dielectric constant.
The strength of the confinement field depends on the specific
fabrication of the boundaries of the 2DES under consideration
(etching, gates, etc.). It has been argued that compressible and
incompressible strips can even become comparable in size \cite{Chkl1}.
Interaction-dominated edge channels are useful in the theory
of the fractional quantum Hall effect \cite{Chan1,Been1,Chkl2,Kane1}.
However, the structure of fractional edge channels is much more complicated,
and we shall restrict our considerations to the integer quantum Hall
regime.\\ \indent 

In the following considerations, three quantities which characterize
the equilibrium state of a sample are important in order to discuss
low frequency transport close to equilibrium.
First, the density of states, $dN_{k}/dE$, of the edge channel $k$ at the
Fermi level gives the
change in the number of states if the electro-chemical potential is
varied for fixed electrostatic potential (i.e. fixed band bottom).
For non-interacting electrons this density of states is determined
by the (equilibrium) velocity of carriers $v(s) = (dU_{eq}(s)/dx)/B$
along the 
the path $s$ of the edge channel, where $x$ is now the transverse
coordinate. It is given by $dN_{k}/dE = \int ds/hv $, where the
integral over $s$ is along the entire path of the edge channel 
from one sample contact to the other. For the interacting model
this density of states diverges at $kT = 0$ since the single particle
potential is flat. But the density of states is finite for any 
non-vanishing temperature. A finite DOS of edge channels due to a
considerable suppression of screening at small
temperatures is indicated by the numerical results presented by
Lier and Gerhardts\cite{Lier1}.\\ \indent
Secondly, one can attribute to the arrangement of metal-like edge
channels a geometrical capacitance matrix $c_{jk}$.
For given geometry, this matrix
can in principle be derived with the help of Poisson's equation.
For metallic screening this capacitance matrix is determined by the
width and location of the edge channels. It is, therefore,
also a function of the magnetic field and the electro-chemical
potentials applied to the contacts and the gates \cite{Govo1}.
\\ \indent
We finally take into account that each edge channel connects
reservoirs in a directed way, due to the uni-directional velocity of
the carriers. This connection is determined by the transmission and 
reflection probability of the contact. In the following,
we shall always regard the just mentioned characteristics of the 
equilibrium state to be given. 
\section{The Nonequilibrium Steady State}
\label{Nonequilibrium}
\subsection{Electro-chemical capacitance of edge channels}
Consider the two-terminal bar of Fig. \ref{fig1}.a under
nonequilibrium conditions. A cross section of the single-particle potential
in the nonequilibrium case is shown in Fig.
\ref{fig2}.b. A small increase of the voltage
$\delta V_{\beta}$ at contact $\beta $, say $\beta =1$,
implies an electro-chemical voltage shift $\delta V_{k} $ in 
channel $k$.
If the transmission probability from the contact into the edge
channel is $1$, then the chemical potential shift of that edge channel
is the same as that of the reservoir 
$\delta V_{k} = \delta V_{\beta}.$
As a consequence, into the edge channel a charge is injected 
which is proportional to the
density of states (DOS) $dN_{k}/dE$ of the edge channel $k$.
This added charge 
creates in the whole sample an electric nonequilibrium  potential
which shifts the band bottom. This leads, in turn, to the injection
of screening charge. The total charge $\delta q_{k}$ in edge channel
$k$ is then given by 
\begin{equation}
\delta q_{k}=D_{k}(\delta V_{k}-\delta U_{k })\;\;,
\label{eq5} 
\end{equation}
where $D_{k}=e^{2}dN_{k}/dE$ is the quantum capacitance of edge channel $k$.  
The nonequilibrium electric potential
$\delta U_{k}$ of edge channel $k$ can be calculated for a given charge
distribution by solving the electrostatic boundary-value problem
associated with Poisson's equation.
This leads to the introduction of the geometrical capacitance matrix
$  c_{kj}$ of the edge-channel configuration by $ \delta q_{k}
=  \sum _{j} c_{k j} \delta U_{j}$. Note, that the $c_{kj}$ are
calculated for edge channels which are {\em disconnected } from
the contacts and where charge is not conserved.
But the relevant (gauge invariant)
potentials are the electro-chemical potentials and not the
electrostatic potentials of disconnected edge channels.
We define thus an
electro-chemical capacitance matrix $ c_{\mu,kj}$ by \cite{Buet7}
\begin{equation}
\delta q_{k}=  \sum _{j} c_{\mu,k j } \; \delta V_{j}  
\label{eq7}
\end{equation}
considering charge conservation, $\sum \delta q _{i} = 0$,
which is due to the connection to the reservoirs.
One finds then from Eqs. (\ref{eq5}) and (\ref{eq7})
$c_{\mu,11}=c_{\mu,22}=
-c_{\mu,12}=-c_{\mu,21}\equiv c_{\mu}$, where the relative electro-chemical
capacitance $c_{\mu}$ of the two edge channels is given by
\begin{equation}
c_{\mu}^{-1}=c_{0}^{-1}+D_{1}^{-1}+D_{2}^{-1} \;\;.
\label{eq8}
\end{equation}
This describes the relative geometrical capacitance
\cite{Landau}, $c_{0}= (c_{11}c_{22}-c_{12}^{2})/(c_{11}+2c_{12}
+c_{22}) $, in series with the quantum capacitances $D_{k}$.\\ \indent
As an example, we consider the non-interacting case where the
widths $\xi _{k}$ of the edge channels $k=1,2$ of length $L_{y}$ are
very small (i.e. $\xi _{k} \approx l_{m}$). For the sake of
simplicity, we assume them to be equal to each other,
$\xi _{k} \equiv \xi $. The distance between the edge channels
is denoted by $L_{x}$, and the charge is to be uniformly
distributed in the edge channels. For line charges,
the geometrical capacitance becomes \cite{Hira1}
$c_{0}=
(L_{y}\pi\epsilon _{0} \epsilon _{r})/(1+\ln(L_{x}/\xi))$.
The density of states $D_{k}$, on the other hand, are given by
$D_{k}=L_{y}e/(2\pi l_{m}^{2}\:|dU_{k}/dx|)$,
where we assumed 
very steep confinement potentials $dU_{k}/dx=(dU/dx)_{x_{k}}$ at the
edge channels located at $x_{k}$. The electro-chemical capacitance
$c_{\mu}$ can then be written as
\begin{equation}
c_{\mu} =\frac{\pi \epsilon _{0} \epsilon _{r}L_{y}}
{1+ \ln (L_{x}/\xi) +\pi (\alpha _{1}+ \alpha _{2})/2}\;\;.
\label{eq9}
\end{equation} 
Here $\alpha _{k}=|dU_{k}/dx| 4 \pi \epsilon _{0} \epsilon _{r}
l_{m}^{2} /e $
is the ratio between the confinement field and the interaction field
at the edge channel $k$.
To be consistent with the non-interacting case, one must have $\alpha
_{k}\gg 1$. Note that $c_{\mu}$
depends on the magnetic field via $\alpha _{k} \propto 1/B  $, and via the
$B$-dependence of the geometry of the edge-channel arrangement.\\ \indent     
Charge conservation in the sample is reflected by the sum rule
\begin{equation}
\sum _{k}c_{\mu,kl}=\sum_{l}c_{\mu,kl}=0\;\;,
\label{eq10}
\end{equation}
which is a well-known property of a set of capacitors where ground
is included. One concludes that two-terminal systems are 
particularly simple since $2\times 2$-matrices satisfying Eq. 
(\ref{eq10}) are characterized
by a single quantity and have thus purely scalar properties.
We will see later on that equations analogous to Eq. (\ref{eq10})
hold also for the dc-conductance and the emittance
\cite{Buet5,Chen1}. Below it will be important that
the electro-chemical capacitance matrix is
symmetric and an even
function of the magnetic field, i.e. $c_{\mu,kl}(B)=c_{\mu,lk}(B)$
and $c_{\mu,kl}(B)=c_{\mu,kl}(-B)$, respectively. These properties are 
evident from our definition of $c_{\mu,kl}$.\\ \indent
\subsection{DC-conductance for a two-terminal Hall bar}
To find the dc-conductance in the transmission approach the current
can be evaluated in response to a small variation 
of the chemical potential of the contacts keeping
the electro-static potential fixed at its equilibrium value. The transmission
probabilities are a functional of the equilibrium electrostatic potential
only. Here we briefly discuss the derivation of the dc-conductance
using the actual non-equilibrium current. For a detailed discussion of the
various possible definitions of currents and their physical interpretation
we refer the reader to Komiyama and Hirai\cite{Komi1}. 
In order to find the dc-conductance $G_{\alpha \beta}^{(0)}$
of the bar in Fig. \ref{fig1}.a, we remark that the
total nonequilibrium current through a contact consists of
two contributions \cite{Thou1,Shik1}. At contact $1$, for example,
a first part $\delta I^{(u)}_{1}\equiv
(e^{2}/h)(\delta U_{1}-\delta U_{2})$ originates from the action of
the nonequilibrium electric field on the occupied equilibrium states
in the Landau level. This part is obtained from a spatial integration
of the current density (\ref{eq3}) in the region between the edge channels.
A second part $
\delta I^{(q)}_{1} \equiv (\delta q_{1} v_{1}+\delta q_{2}
v_{2})/L_{y}$
is caused by the motion of the added charge density $\delta q_{k}/L_{y}$
with an equilibrium velocity $v_{k}$ in edge channel $k$. 
In the present notation, the relation between the velocity and the
DOS for quasi one-dimensional conductors reads 
$ v_{k}=\pm L_{y}e^{2}/(hD_{k}) $ where the sign is different for
opposite edge channels. Using this and Eq. (\ref{eq5}) gives 
$\delta I^{(q)}_{1} = (e^{2}/h) (\delta V_{1}-\delta U_{1}) -
(e^{2}/h) (\delta V_{2}-\delta U_{2})$. It follows immediately
that the total current $\delta I_{1} = 
\delta I^{(u)}_{1}+\delta I^{(q)}_{1}$ depends only on 
the electro-chemical potentials of the contacts and is given by
$\delta I_{1} = (e^{2}/h)(\delta V_{1}- \delta V_{2})$ with a 
zero-frequency conductance $G^{(0)}\equiv
G_{11}^{(0)}=G_{22}^{(0)}=-G_{12}^{(0)}=-G_{21}^{(0)}=e^{2}/h$.
This universal result reflects the integer quantum Hall effect
\cite{Klit1}. Using  $\delta q_{1} = c_{0} (\delta U_{1}-\delta U_{2})$ and
$\delta q_{1}= c_{\mu} (\delta V_{1}-\delta V_{2})$ with $c_{\mu}$ given by  
Eq. (\ref{eq8}) we find for the ratio of the two currents
$\delta I^{(q)}/\delta I^{(u)}=(D_{1}^{-1}+D_{2}^{-1})\: c_{0}$. 
This ratio is large for small DOS $D_{k}$ of the edge channels, i.e.
for a sufficiently steep slope of the confinement potential, when
the chemical contribution to the current predominates.
On the other hand, if the edge channels
are \em macroscopic \rm metallic conductors with complete screening,
($\delta U_{\alpha}\to \delta V_{\alpha} $ and
$v_{k}\to 0$) the chemical contribution vanishes and the electrostatic
contribution predominates.
\subsection{DC-conductance for an $M$-terminal sample with $N$ edge channels} 
Consider now a more general quantum Hall sample with $M$ contacts and
$N$ edge channels. 
We assume that the density of states, $D_{k}$,
and the electro-chemical
capacitance matrix, $ c_{\mu, kj}$ ($k,j=1,...,N$), are known.
An expression for $c_{\mu,kj}$ in terms of the geometrical
capacitance, $c_{kj}$, and the DOS of the edge channels, $D_{k}$,
is derived
in an appendix. Eqs. (\ref{eq5}) and (\ref{eq7}) are still valid
in the present case.
Each edge channel $k$ is connected to reservoirs $\alpha $ and $\beta $, where
$\alpha = \beta $ is permitted. Reservoir $\beta $
injects carriers into edge channel $k$ from which
carriers are emitted into reservoir $\alpha $. 
For simplicity and to be definite,
we assume that the contact resistances of this sample
are quantized\cite{vanW2,Muel1}. The transmission probability of a
carrier in contact $\beta$ to enter edge channel $k$ is denoted by
$\Delta _{k \beta}(B)$ and for quantized contact resistances 
is given by  
\begin{equation}
\Delta _{k\beta}(B) =\cases{1 &${\rm if \; contact } \;\; \beta
{\rm \;\; injects \; into \; channel \;\; } k$ \cr 
                          0 &$\;\;\;\; {\rm otherwise} $ \cr} \;\;.
\label{eq16}
\end{equation}
Similarly, we introduce the probability of a carrier which approaches
contact $\beta$ on an edge channel $k$ to enter the contact
\begin{equation}
\Delta _{\alpha l}(B) =\cases{1 &${\rm if \; channel} \;\; l
{\rm \;\; emits \; into \; contact \;\; } \alpha $ \cr 
                          0 &$\;\;\;\; {\rm otherwise} $ \cr} \;\;.
\label{eq16b}
\end{equation}
From the micro-reversibility properties of the transmission probabilities
we have  $\Delta _{\beta k}(B)= \Delta _{k \beta}(-B)$. 
The transmission probability of the contact plays the
role of a topological factor determined by the connectivity of the 
edge channel to the contacts of the sample.
With the help of the contact transmission probability (\ref{eq16}), the
variation $\delta V_{k}$ of the electro-chemical voltage of edge channel
$k$ can be expressed in terms of the voltages $\delta V_{\beta}$
in the reservoirs:
\begin{equation}
\delta V_{k}=\sum _{\beta=1}^{M} \Delta _{k\beta}\: \delta
V_{\beta} \;\;.
\label{eq16a}
\end{equation}
The charge $\delta q _{k}$ in edge channel $k$ is then related to voltage
variations in the contacts by
\begin{equation}
\delta q_{k}=  \sum _{\beta=1}^{M} \sum _{l=1}^{N}
c_{\mu,k l } \Delta _{l\beta}  \; \delta
V_{\beta} \;\;. 
\label{eq17}
\end{equation}
The total charge in all those channels into which contact $\alpha $
injects is $\sum_{k}\delta
q_{k}\: \Delta _{k\alpha} $. If there is no transmission (i.e. if all edge
channels are connected to a single contact), one has
$\Delta _{k\alpha } = \Delta _{\alpha k}$.  
A capacitance measurement yields then a capacitance matrix
\begin{equation}
C_{\mu ,\alpha \beta }=  \sum_{k,l=1}^{N} 
\Delta _{k \alpha }c_{\mu,kl} \Delta _{l\beta} \;\;.
\label{eq18}
\end{equation}
Below, we shall see that the assumption of the absence of
transmission between different contacts is crucial in order to find a
magneto-capacitance according to Eq. (\ref{eq18}).\\ \indent 
The dc-conductance $\delta I_{\alpha}/\delta V_{\beta} $
can be calculated following the same lines as above for the
two-terminal case with a single Landau level.
The current $\delta I_{\alpha}$ through contact
$\alpha $ is obtained from a sum over
all incoming and outgoing channels $k$ with a
contribution $\delta
I^{(q)}_{k}$ and over all Landau levels with a
contribution $\delta I^{(u)}_{k}$. The well-known result \cite{Buet1}
reads in our notation  
\begin{equation}
G_{\alpha\beta }^{(0)} = \frac{e^{2}}{h} \biggl( K_{\beta}\delta _{\alpha
\beta} - \sum _{k} \Delta _{\alpha k} \Delta _{k\beta} \biggr)
\;\;. 
\label{eq19}
\end{equation}
For the derivation of Eq. (\ref{eq19}) we used that
$\sum _{k}\Delta _{k\beta}\: \Delta _{k\alpha}\:
=K_{\beta}\delta _{\alpha \beta}$ where  $K_{\beta}$ is the
number of edge channels in which contact $\beta $ injects.
The diagonal element, $G_{\beta \beta}^{(0)}$, is $e^{2}/h$ times
the number of channels
which leave contact $\beta $ and which do not return to this contact,
and $-G_{\alpha \beta}^{(0)}$ ($\alpha \neq \beta $) is $e^{2}/h$ times
the number of directed channels going from contact $\beta $ to contact
$\alpha$. Note that both the current conservation property 
$\sum _{\alpha} G_{\alpha \beta}^{(0)}=\sum_{\beta}G_{\alpha
\beta}^{(0)}=0$ 
and the Onsager-Casimir reciprocity relations, $G_{\alpha
\beta}^{(0)}(-B)=G_{\beta \alpha}^{(0)}(B)$, are satisfied.\\ \indent
\section{The Emittance Matrix} 
\label{Emittance}
\subsection{Emittance matrix for general mesoscopic conductors}
First, we recall the theory of the emittance
for a general arrangement
of mesoscopic conductors by closely following Ref. \cite{Buet5}.
Once the electrostatic geometrical capacitance matrix is known, 
we can formulate our discussion in terms of a discrete set of 
potentials which we take to be constants along each edge channel.
A general formulation of the theory for such a discrete potential
model is the subject of Ref.\cite{Pret1}.  
It is well-known that the transmission approach to
current transport relates conductances to
scattering matrices of the conductors. 
A scattering matrix relates incoming and outgoing current amplitudes
of the contacts $\alpha = 1, ... , M$ of a
sample for each conduction channel $k=1, ... , N$.
The DOS $dN_{k}/dE$ of channel $k$ expressed in terms of
the scattering matrix can then be written as a sum of
\em partial densities of states, \rm $dN_{\alpha k \beta}/dE$.
The quantity $dN_{\alpha k \beta}/dE$ is the DOS of channel $k$
associated with carriers which are scattered from contact $\beta $ to contact
$\alpha $. A slight variation $\delta V_{\beta}$ of the voltage 
in contact $\beta $ causes the injection of a total charge $\delta Q
_{\alpha}$ through contact $\alpha $. Thus, a slowly oscillating voltage
implies an additional current 
$  -i\omega \delta Q_{\alpha}\exp(-i\omega t) $
at this contact. Now, it follows from the definition in Eq. (\ref{eq2})
that the emittance $E_{\alpha \beta }$ can be identified with
$\delta Q_{\alpha}/\delta V_{\beta}$. Note that there are two contributions
to $\delta Q_{\alpha}$. A first part which neglects
screening is given by a kinetic contribution
\begin{equation}
\delta Q^{(k)} _{\alpha} = e^{2}\sum _{k=1}^{N}\sum _{\beta =1} ^{M}
\frac{dN_{\alpha k\beta}}{dE} \; \delta V_{\beta }\;\;.
\label{eq20}
\end{equation}
This part gives the charge which is scattered from the contacts
$\beta $ to contact $\alpha $ due to the shift of the electro-chemical
potentials $\delta V_{\beta}$ at fixed electrostatic potentials
$\delta U_{k}$. However, the nonequilibrium electrostatic
potential which is due to the nonequilibrium charge-distribution in 
the edge channels is still neglected in Eq. (\ref{eq20}). In order to 
take it into account, we recall that the $\delta U_{k}$ are shifts of the 
band bottoms of the edge channels, which 
cause an induction of additional screening charges. Hence, there is a second contribution $\delta Q ^{(s)}_{\alpha}$
given by the part of the screening charge which is eventually scattered
to contact $\alpha $. It can be expressed in the form
\begin{equation}
\delta Q ^{(s)}_{\alpha} = - e^{2}
\sum _{k=1}^{N} \left( \sum _{\beta =1}^{M}
\frac{dN_{\alpha k\beta}}{dE}\right) \delta U_{k}
\;\;.
\label{eq21}
\end{equation}
The quantity in the large bracket, $
dN_{\alpha k}/dE \equiv   \sum _{\beta=1}^{M} dN_{\alpha k \beta
}/dE$, is a partial DOS associated with carriers in channel $k$
emitted into contact $\alpha $ irrespective of the injecting contact
$\beta $. The change
of the electric potential $\delta U_{k}$ at channel $k$ is
determined by the variations of the electro-chemical
potentials of the conductors. Within linear response theory
we write
\begin{equation}
\delta U_{k} = \sum_{\beta = 1}^{M}
u_{k\beta} \; \delta V_{\beta}\;\;,
\label{eq23}
\end{equation}
where the \em characteristic potentials \rm $u_{k\beta}$ \cite{Buet5}
give the change of the electrostatic potential of conductor
$k$ if the voltage is changed in contact $\beta $ by unity.
The sum of the two parts given by Eqs. (\ref{eq20}) and (\ref{eq21})  
leads finally to the emittance matrix $\delta Q_{\alpha}/\delta
V_{\beta} $:
\begin{equation}
E_{\alpha \beta }= e^{2} \sum _{k=1}^{N}\biggl(\frac{dN_{\alpha k
\beta}}{dE} -
\frac{dN_{\alpha k}}{dE}u_{k\beta}\biggr) \;\;.
\label{eq24}
\end{equation}
The occurrence of the characteristic potentials $u_{k\beta }$
indicates the
necessity of the knowledge of the nonequilibrium state in order
get the emittance. Since the characteristic potentials are sample
specific, one cannot expect to obtain a
universal result for the ac-admittance. Furthermore, since the kinetic part
and the screening part contribute with opposite signs,
the emittance elements can have positive or negative sign
depending on which part is dominant \cite{Buet5}.
\subsection{Emittance matrix for quantized Hall samples}
To apply the result (\ref{eq24}) to quantized Hall samples, one
uses the fact that the partial DOS can be expressed in terms of
the transmission probabilities $\Delta _{\alpha l}$,
$\Delta _{k \beta }$, and the 
density of states $D_{k}$ of edge channel $k$:
\begin{equation}
e^{2}\frac{dN_{\alpha k\beta}}{dE}=
\Delta_{\alpha k} \: D_{k}\: \Delta_{k\beta } \;\;\;\; , \;\;\; 
e^{2} \frac{dN_{\alpha k }}{dE}= \Delta_{\alpha k} \: D_{k}\;\;.
\label{eq25}
\end{equation}
This follows directly from the suppression of backscattering in an 
edge channel. The characteristic potentials $u_{k \beta}$ follow from
Eqs. (\ref{eq5}), (\ref{eq7})
and (\ref{eq16a}):  
\begin{equation}
u_{k \beta}=\sum_{l=1}^{N} (\delta _{kl}-
D_{k}^{-1} c_{\mu, kl})\;\Delta
_{l\beta}\;\;.
\label{eq26}
\end{equation}
By inserting Eqs. (\ref{eq25}) and (\ref{eq26}) in Eq. (\ref{eq24})
one obtains 
\begin{equation}
E_{\alpha \beta}= \sum_{k,l=1}^{N} \Delta _{\alpha k} \:
c_{\mu,kl} \: \Delta _{l\beta}\;\;.
\label{eq27}
\end{equation}
This is the key result of our work. The emittance is the 
sum of all those charges which are \em emitted \rm at contact $\alpha $
due to the injection of charge at contact $\beta $ mediated
by Coulomb interaction between edge channels. The elementary process
which contributes to the emittance is illustrated
by the diagram in Fig. \ref{fig3}.\\ \indent
The emittance has the following properties.
From $\sum _{\alpha} \Delta _{k\alpha} =1$ and
Eq. (\ref{eq10}) one concludes that 
\begin{equation}
\sum_{\beta}E_{\alpha \beta}=\sum_{\alpha}E_{\alpha \beta }\equiv 0
\label{eq28}
\end{equation}
which is a consequence of charge neutrality\cite{Buet5}.
Since $\Delta _{\alpha k}(-B) = \Delta _{k \alpha }(B)$,
the Onsager-Casimir reciprocity relations\cite{Buet5} 
\begin{equation}
E_{\alpha \beta}(B)=E_{\beta \alpha}(-B),
\label{eq29}
\end{equation}
based on micro-reversibility are satisfied, too.
In contrast to $c_{\mu, kl}(B)$, the emittance matrix
$E_{\alpha\beta}(B)$
is in general {\em not} symmetric. A comparison of the
Eqs. (\ref{eq18}) and (\ref{eq27}) implies that the emittance is a
(symmetric) capacitance, i.e.
$E_{\alpha \beta}\equiv C_{\mu, \alpha \beta}$,
if each edge channel $k$ is connected to a single
reservoir, i.e. if $\Delta _{k\alpha}(B) \equiv \Delta _{\alpha k}(B) $
holds.\\
\indent
\section{Examples}
\label{Examples}
In this section we apply the previous results to various examples
of Hall devices. The electro-chemical capacitance matrix $c_{\mu ,kj}$
is always assumed to be known. 
\subsection{Two-terminal devices}
The two-terminal devices in Figs. \ref{fig1} and \ref{fig4}.a
can be characterized by
the scalar admittance $G= G^{(0)}-i\omega E \equiv
G_{11}=G_{22}=-G_{12}=-G_{21}$. While $G^{(0)}=e^{2}/h$ for the Hall
bar in Fig. \ref{fig1}.a, the dc-conductance vanishes identically for the
Corbino disc in Fig. \ref{fig1}.b since there is no dc-current
flowing through the contacts. From Eq. (\ref{eq27}) one finds
the emittances $E= -c_{\mu}$ and $E= c_{\mu}$
for the bar and the disc, respectively. Here, $c_{\mu}$ denotes the
relative electro-chemical capacitance between the edge channels. 
While the emittance of a Corbino disc is an electro-chemical
capacitance, the emittance of a quantum Hall bar turns out to be
a {\em negative} electro-chemical capacitance.
The interchange of the sign can be understood intuitively by remarking that
the kinetic part $\delta Q^{(k)}$ and the part $\delta Q^{(s)}$
associated with screening are interchanged for the two
different topologies. Indeed, for a voltage oscillation at contact $1$
of the bar, transmitted charge goes to reservoir $2$ and induced charge
comes back via edge channel $2$. In the Corbino geometry, on the other hand,
transmitted charge comes back to contact $1$ and screening charge
goes to reservoir $2$.\\ \indent

In order to obtain an intuition for the signs of emittances,
consider Fig. \ref{fig4}.a where a two-terminal
quantum Hall bar with two pairs of edge channels is shown. 
A constriction is assumed to bend back
the second pair (3 and 4) which will thus not
contribute to the dc-conductance. The dc-conductance is
$e^{2}/h$ as for the case of the
bar in Fig. \ref{fig1}.a. However, the second pair gives a capacitive
contribution to a time dependent current.   
From Eq. (\ref{eq27}) and using Eq. (\ref{eq10}), one finds
immediately $E= c_{\mu ,12}-c_{\mu ,34}$. Hence, the emittance is a
capacitance (i.e. $E>0$) if the Coulomb interaction between edge 
channels 
$3$ and $4$ is stronger than the Coulomb interaction between edge 
channels $1$ and
$2$. The transmitting edge channels contribute thus with a negative
capacitance.
It is very remarkable that only two elements of
the full capacitance matrix $c_{\mu ,kl}$ determine the emittance.
This is a consequence of the quantized contact transmission-probabilities
and of the symmetry and current-conservation properties 
of the capacitance matrix. The direct way in which our approach permits
to derive this result demonstrates its usefulness.
\\ \indent

\subsection{Three-terminal device: a bar with additional gate}
The three-terminal device in Fig. \ref{fig4}.b consists of a quantum
Hall bar with a gate on top of the 2DES and close to one sample
edge. The gate is connected to a further contact and couples only
capacitively to the edge channels. This setup has been investigated in 
Ref. \cite{Chen1}. Clearly, all the $G_{\alpha 3}$ and
$G_{3\beta }$ vanish. The dc-conductance for the contacts
$1$ and $2$ is equal to $G_{\alpha \beta}^{(0)}$ for the
quantum Hall bar in Fig. \ref{fig1}.a. The presence of the gate breaks the
symmetry of the quantum Hall bar where at
equal time the magnetic field $B$ is inverted and the reservoirs
$1$ and $2$ are interchanged. One
expects thus that the emittance matrix
${\bf E}$ is an asymmetric function of the magnetic field.
Equation (\ref{eq27}) yields
\begin{equation}
{\bf E}(B) = \left( \begin{array}{ccc}
c_{\mu,21 }& c_{\mu,22 } & c_{\mu,23 }\\
c_{\mu,11 }& c_{\mu,12 } & c_{\mu,13 }\\
c_{\mu,31 }& c_{\mu,32 } & c_{\mu,33 }
\end{array} \right)\;\;.
\label{eq29a}
\end{equation} 
For instance, by measuring the current at contact $1$ for a voltage
oscillation at the gate, one finds $E_{13}(B)= c_{\mu,23}$ for one
polarity of the field $B$, 
but $E_{13}(-B)= c_{\mu,31}$ for the other field polarity. This follows
directly from the reciprocity relations (\ref{eq29}).
Because the capacitance $c_{\mu,13}$
between channels $1$ and $3$ is different from the capacitance
$c_{\mu,23}$ between channels $2$ and $3$, one observes a completely
asymmetric
emittance coefficient $E_{13}(B)$ as a function of the magnetic field. 
This prediction is in agreement with the experimental
results reported in Ref. \cite{Chen1}.
The symmetry of the emittance matrix strongly reflects
the geometry of the edge-channel arrangement.\\ \indent
\subsection{Four-terminal Hall bars}
In Fig. \ref{fig5}, four-terminal samples
are shown which are used in order to investigate
the quantum Hall effect \cite{Klit1}.
In such devices, two contacts serve as current source
and sink, whereas the two remaining contacts are used as voltage
probes. In Fig. \ref{fig5}.a an ideal bar is shown where edge channels 
connect subsequent contacts along the sample edge. In the sample
of Fig. \ref{fig5}.b, on the other hand,
there are certain edge channels leaving one and the same contact but
connecting different contacts.\\ \indent 
Let us assume for the ideal four-terminal bar in Fig. \ref{fig5}.a
a filling factor between the integers $p$ and $p+1$
such that
$p$ edge channels exist along each sample edge which connect
contact $k$ with contact $k+1$. It is possible to
define electro-chemical capacitances $c_{\mu ,jk}$ between these
sets of edge channels which leave contact $k$ and of those which
leave contact $j$. For each of those sets we 
plotted a single directed line. The Eqs. (\ref{eq19})
and (\ref{eq27}) yield $G_{\alpha \beta}^{(0)}=g\:
(\delta _{\alpha \beta } -\delta _{\alpha -1 \:\beta })$
for the dc-conductance, and $E_{\alpha \beta}=c_{\mu,
\alpha-1\:\beta}$ for the
emittance, respectively. Here, we defined $g=p(e^{2}/h)$, and
the indices $0$ and $4$ have to be identified
with each other.\\ \indent
On the other hand, for the specific non-ideal Hall bar plotted in
Fig. \ref{fig5}.b the connection between contacts via edge channels
is not simply determined by the topology of the sample boundary.
In the particular case of Fig. \ref{fig5}.b, the dc-conductance becomes
\begin{equation}
{\bf  G}^{(0)}= \frac{e^{2}}{h}\left( \begin{array}{cccc}
  2  &  0  &  -1  & -1 \\
 -1  &  1  &  0  & 0 \\
  -1  & -1  &  2  &  0 \\
  0  & 0  & -1  &  1
\end{array} \right) \;\;.
\label{eq29b}
\end{equation} 
For the emittance coefficients one finds expressions of the form
$E_{11}= c_{\mu ,14}
+c_{\mu ,17}+c_{\mu ,54}+c_{\mu ,57}$ etc., where the $c_{\mu ,kl}$
denote the electro-chemical capacitances between edge channels
labelled as shown in Fig. \ref{fig5}.b.
Below, we will use this example in order to discuss the effect of
voltage probes. Furthermore, we will derive the frequency dependent
longitudinal and Hall resistances for the ideal Hall sample
in Fig. \ref{fig5}.a.\\ \indent

\section{Effect of voltage probes}
\label{incoherent}
In this section we study the crossover from a $M$-terminal sample
to a $M-1$-terminal sample by using one contact, say
contact $\Omega$, as a voltage probe. For the dc-conductance,
this problem has been investigated in Ref.\cite{Buet8}.
We assume that there is at least one edge
channel which connects contact $\Omega $ with a different contact.
For an ideal voltage probe, there is no possibility for charge to
pass through contact $\Omega$ such that $\delta I_{\Omega }\equiv 0$.
By eliminating $\delta V_{\Omega} $ in Eq. (\ref{eq1}), one obtains
from
\begin{equation}
\delta V_{\Omega} = - \frac{1}{G_{\Omega \Omega}^{(0)}}
\sum_{\beta \neq  \Omega} \left(
G_{\Omega \beta} ^{(0)} -i\omega (E_{\Omega \beta}- 
\frac{G_{\Omega \beta}^{(0)}}{G_{\Omega \Omega}^{(0)}} E_{\Omega
\Omega}) \right)\: \delta V_{\beta}
\label{eq31a} 
\end{equation}
a new admittance $\tilde G_{\alpha \beta}
(\omega) =\tilde G_{\alpha \beta}^{(0)}-i \omega \tilde E_{\alpha \beta}$
for the remaining $M-1$ contacts, where
\begin{eqnarray}
\tilde G_{\alpha \beta}^{(0)} & = & G_{\alpha \beta}^{(0)}
- \frac{G_{\alpha \Omega}^{(0)}G_{\Omega \beta}^{(0)}}
{G_{\Omega \Omega}^{(0)}}
\label{eq30}\\
\tilde E_{\alpha \beta} & = & E_{\alpha \beta}
+\frac{G_{\alpha \Omega}^{(0)}G_{\Omega \beta}^{(0)}}
{(G_{\Omega \Omega}^{(0)})^{2}}E_{\Omega \Omega}
-\frac{G_{\alpha \Omega}^{(0)}}{G_{\Omega \Omega}^{(0)}}
E_{\Omega \beta}
-E_{\alpha \Omega} 
\frac{G_{\Omega \beta}^{(0)}}{G_{\Omega \Omega}^{(0)}} \;\;.
\label{eq31}
\end{eqnarray} 
A brief calculation confirms that Eqs. (\ref{eq28}) and (\ref{eq29})
remain valid for Eqs. (\ref{eq30}) and (\ref{eq31}).
The additional terms appearing in Eqs. (\ref{eq30}) and (\ref{eq31}) describe 
scattering between edge channels at contact $\Omega $ 
(incoherent terms) \cite{Buet8}.
Now, the probability of a carrier to go from contact $\beta $ to
contact $\alpha $ is no longer restricted to the values zero and unity.
The additional terms have the following
simple interpretations. Firstly, the second term on the right hand side
of Eq. (\ref{eq30}) describes the equipartition of
the current which comes from contact $\beta $ to $\Omega $
between the channels which go from contact $\Omega $ to $\alpha $.  
Secondly, the three correction terms on the right hand side of
Eq. (\ref{eq31}) can be associated with
processes \\
1) where carriers go from contact $\beta $ to
contact $\alpha $ bypassing $\Omega $ and obtain a
`self-emittance' contribution $E_{\Omega \Omega}$,\\ 
2) where carriers induced
via the emittance $E_{\Omega \beta}$ are transmitted from
contact $\Omega $ to contact $\alpha $, and \\
3) where carriers which are transmitted from contact $\beta $ to
contact $\Omega$ interact via the emittance $E_{\alpha \Omega}$
with contact $\alpha $.\\ \indent

As an example, we consider the four-terminal sample of Fig.
\ref{fig5}.b where contact $3$ is to serve as the voltage probe.
The Eqs. (\ref{eq30}) and (\ref{eq31})
yield the following three-terminal
dc-conductance for the contacts $1$, $2$, and $4$:
\begin{equation}
{\bf \tilde G}^{(0)}= \frac{e^{2}}{h}\left( \begin{array}{ccc}
3/2& -1/2 & -1\\
-1 & 1 &  0 \\
-1/2 & -1/2 & 1
\end{array} \right)\;\;.
\label{eq32}
\end{equation} 
For example, carriers from edge channel $2$ 
will be scattered at contact $3$ with probability one half
to channel $3$ and one half to channel $7$, which implies
$\tilde G^{(0)}_{42}=\tilde G_{12}^{(0)}=-e^{2}/2h $. Similar
interpretations can be found for the other elements of the
dc-conductance matrix (\ref{eq32}).\\ \indent
From Eq. (\ref{eq31}) one obtains the emittance matrix
\begin{equation}
{\bf \tilde E}= {\bf \hat E} + \left( \begin{array}{cccc}
E_{33}/4 +E_{13}/2+E_{31}/2   & E_{33}/4 +E_{13}/2+E_{32}/2 &
E_{34}/2 \\
E_{23}/2  & E_{23}/2  & 0   \\
E_{33}/4 +E_{43}/2+E_{31}/2  & E_{33}/4 +E_{43}/2+E_{32}/2  & E_{34}/2
\end{array} \right)\;\;.
\label{eq33}
\end{equation} 
where the $3\times 3$-matrix $ {\bf \hat E}$ is obtained from the
matrix ${\bf E}$ by deleting row $3$ and column $3$.  
The fact that $\tilde E_{24}= E_{24}$ holds can be easily understood from
Eq. (\ref{eq31}): there are neither edge channels which go from contact
$4$ to $3$ nor from contact $3$ to $2$. Simple interpretations
exist also for the other emittance coefficients. For example, consider the
additional term $E_{23}/2$ of $\tilde E_{22}$. A voltage
oscillation in contact $2$ induces a current in edge channel
$2$ which leads to contact $3$. This current is
divided into {\em two} parts (channels $3$ and $7$) at contact $3$.
Hence, a contribution $E_{23}$ with a factor one half occurs.
\section{Longitudinal and Hall resistances at low frequencies}
\label{Hall}
The integer quantum Hall effect 
corresponds to the quantization of the Hall resistance and
the vanishing of the longitudinal resistance of the ideal
four-probe quantum Hall bar of Fig. \ref{fig5}.a
at zero frequency \cite{Klit1}. Two of the contacts
serve as voltage probes, whereas the two remaining contacts are used
as source and sink for the current. The discussion of the
quantum Hall effect in terms of edge channels is provided by Ref.
\cite{Buet1}. With the help of the theory
presented in this paper, the results of these references can now be
extended to the low frequency case.\\ \indent
If the contacts $3$ and $4$ in Fig. \ref{fig5}.a are
the voltage probes, the longitudinal resistance is defined
by $R_{L}=R_{12,34}= (\delta V_{3}-\delta V_{4})/ \delta I_{1}$.
On the other hand,
the Hall resistance is defined by $R_{H}=R_{13,24}=
(\delta V_{2}-\delta
V_{4})/\delta I_{1}$, provided the contacts $2$ and $4$ are voltage
probes. With the help of Eq. (\ref{eq1}), $R_{L}$ and $R_{H}$ can be
expressed in terms of the $G_{\alpha  \beta}$. After some linear
algebra one finds \cite{Buet1} $R_{L}=(G_{32}G_{41}-G_{31}G_{42})/D$
and $R_{H}=(G_{21}G_{43}-G_{41}G_{23})/D$, where $D$ is the
determinant of the $3\times 3$ matrix $G_{\alpha \beta}$ restricted
to, say, $\alpha , \beta = 1,...,3 $. By using the results of
Sect. \ref{Examples}.c,
one obtains a longitudinal resistance $R_{L} = i\omega
E_{41}/g^{2}$, where $g=pe^{2}/h$ with $p$ being the number of edge
channels along an edge. With $E_{41}= c_{\mu ,13 }$ one obtains 
\begin{equation}
R_{L} =i\omega \: \frac{c_ {\mu ,13}}{g^{2}} \;\;.
\label{rlong}
\end{equation} 
The leading term of the longitudinal resistance is
determined by the Coulomb coupling between the current
circuit and the voltage circuit which are represented by edge channels
$1$ and $3$, respectively. On the other hand, the 
Hall resistance turns out to be 
\begin{equation}
R_{H}=\frac{1}{g}+ i\omega \: \frac{ c_{\mu,
24}-c_{\mu ,13}}{g^{2}}  \;\;.
\label{rhall}
\end{equation} 
This result implies that,
in contrast to the longitudinal resistance, for the Hall
resistance the sign of
the first-order term with respect to frequency depends on the
specific locations of the contacts. In principle, the
capacitances $c_{\mu ,24}$ and
$c_{\mu ,13}$ can be found independently by measuring
$R_{L}$ for appropriate choices of current and voltage probes.
A further measurement of $R_{H}$ provides then a test for the validity 
of Eqs. (\ref{rlong}) and (\ref{rhall}).
Finally, a direct calculation shows that
the Eqs. (\ref{rlong}) and (\ref{rhall}) satisfy the
reciprocity relations \cite{Buet1} $R_{jk,mn}(-B) = R_{mn,jk}(B)$;
in particular, $R_{L}(-B) = R_{L}(B)$ holds. 
\section{Summary}
We investigate the low frequency admittance of
quantized Hall samples by using a simple discrete potential model
based on the decomposition of the 2DES in (well-separated)
metallic and dielectric parts
and by applying a general theory of the low frequency admittance
$G_{\alpha \beta}=G_{\alpha \beta}^{(0)}-i\omega E_{\alpha \beta}$
for mesoscopic conductors. The main result
is an expression for the emittance matrix $E_{\alpha \beta}$ in
terms of electro-chemical capacitance elements which depend
on the geometrical configuration and the density of states of the edge
channels. We emphasize that the theory satisfies the important
requirement of charge neutrality and current conservation. The
emittance gives the charge emitted through contact $\alpha $
mediated by the Coulomb interaction of edge channels for a
voltage-variation at contact $\beta $.
If there is no transmission of charge between
different reservoirs,
the emittance is a capacitance, but in the presence of transmission
the emittance can even be a negative capacitance. This has been 
exemplified by comparing Corbino and bar geometries.
The symmetry of the
emittance matrix with respect to the magnetic field depends
significantly on the geometry of the edge channels. The presence of a
voltage probe and the resulting inter edge-channel scattering at the 
voltage probe is investigated.
We finally derive expressions for the frequency
dependent longitudinal and Hall resistances of an ideal four-probe bar.
Due to the intuitive expression for the emittance,
all results have simple interpretations.\\ \indent
        
\em Acknowledgement \rm
This work was supported by the Swiss National Science Foundation.
\section*{Appendix}
In order to derive the
electro-chemical capacitance matrix $c_{\mu , kl}$ for
a system with many edge channels,
we assume that the spatial variation of the electrostatic potential
{\em inside} the edge channels can be neglected (discrete potential
model). Then, Eq.
(\ref{eq5}) remains valid and the definition of a geometrical
capacitance $c_{jk}$ is meaningful. It is convenient to use
vector and matrix notation. Let us write the $N\times N$-matrix 
for the geometrical capacitance of the (disconnected) edge channels
by ${\bf c} \equiv c_{jk} $, and in a
similar way for the electro-chemical
capacitance ${\bf c_{\mu}}\equiv c_{\mu, jk}$, and the DOS ${\bf D}\equiv
D_{k}\delta_{jk}$. We introduce $N$-dimensional vectors
${\bf \delta q}$, ${\bf \delta U}$, and ${\bf \delta V}$
for the
charges, the electrostatic and the electro-chemical potentials of the
edge channels, respectively. The solution of the Poisson equation
for a given charge distribution yields an electric potential
${\bf \delta U }={\bf
c}^{-1}{\bf \delta q}+\delta U^{(0)}{\bf 1}$, where ${\bf 1}$ is a
vector with all components being unity. Note that a constant
potential shift $\delta U^{(0)}$ in the whole sample
is always a solution of the
Poisson equation and is determined by charge conservation,       
$\sum_{k}\delta q_{k}=0$. Hence, $\delta U^{(0)}= \sum
_{j,k}c_{jk}\delta U_{k}\;/\;\sum _{j,k}c_{jk}$ which defines
a matrix ${\bf \Lambda }$ such that $\delta U^{(0)}{\bf 1} =
{\bf \Lambda }{\bf \delta U}$. 
The electro-chemical capacitance matrix ${\bf \delta q / \delta V}$
follows from
${\bf \delta q} = {\bf D}({\bf \delta V}
-{\bf \delta U})={\bf c}({\bf \delta U} -\delta U^{(0)} {\bf 1})$
and can be expressed in the form 
\begin{equation}
{\bf c_{\mu}}= ({\bf c}^{-1} + ({\bf I}-{\bf \Lambda})
{\bf D}^{-1} )^{-1} ({\bf I }-{\bf \Lambda})
\label{ap1}
\end{equation} 
where ${\bf I}$ denotes the identity matrix. 
\newpage
\begin{figure}
\caption{{\bf a)} Quantum Hall bar with a single
pair of edge channels (thin directed lines) and connected to two
reservoirs at electro-chemical potentials $\mu _{1,2}$.
{\bf b)} Corbino disc with contacts at the inner and the outer edges.}
\label{fig1}
\end{figure}

\begin{figure}
\caption{{\bf a)} Single particle potential for a transverse cross
section of the Hall bar in Fig. 1.a. Empty, partially filled, and
filled circles correspond to empty, partially filled, and filled
states, respectively. Edge channels are the partially filled and 
extended  
states at the Fermi energy $E_{F}$ close to the sample edge, where the 
potential is dominated by the confinement potential.
{\bf b)} Nonequilibrium version of part a. The electro-chemical
voltage variation
$\delta V_{1}$ induces charges $\delta q_{1,2}$
and nonequilibrium electrostatic
potential shifts $\delta U_{1,2}$ in the edge channels.}
\label{fig2}
\end{figure}

\begin{figure}
\caption{ An emittance element $E_{\alpha \beta} $ is the sum
over all electro-chemical capacitance elements $c_{\mu ,kl}$
of the edge channels $k$ and $l$ which correspond to the elementary 
process shown in this figure.}
\label{fig3}
\end{figure}

\begin{figure}
\caption{\rm {\bf a)} Two-terminal bar with constriction. Only one 
pair of edge channels connects different reservoirs, whereas
another pair returns to the original reservoir. 
{\bf b)} Three-terminal Hall bar with gate ($3$).
The asymmetric geometry leads to an asymmetric magneto-capacitance
as a function of the magnetic field.}
\label{fig4}
\end{figure}

\begin{figure}
\caption{{\bf a)} Ideal four-terminal Hall bar. The geometry of edge
channels is determined by the sample boundary.
{\bf b)} Four-terminal Hall bar with complicated edge-channel 
arrangement.}
\label{fig5}
\end{figure}

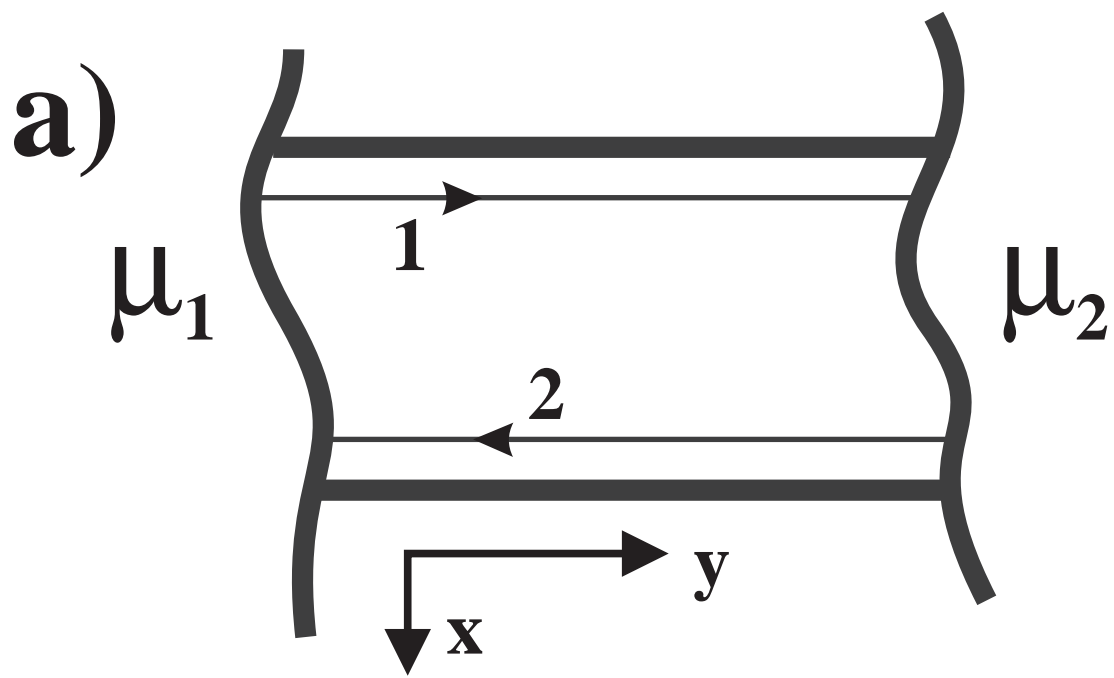
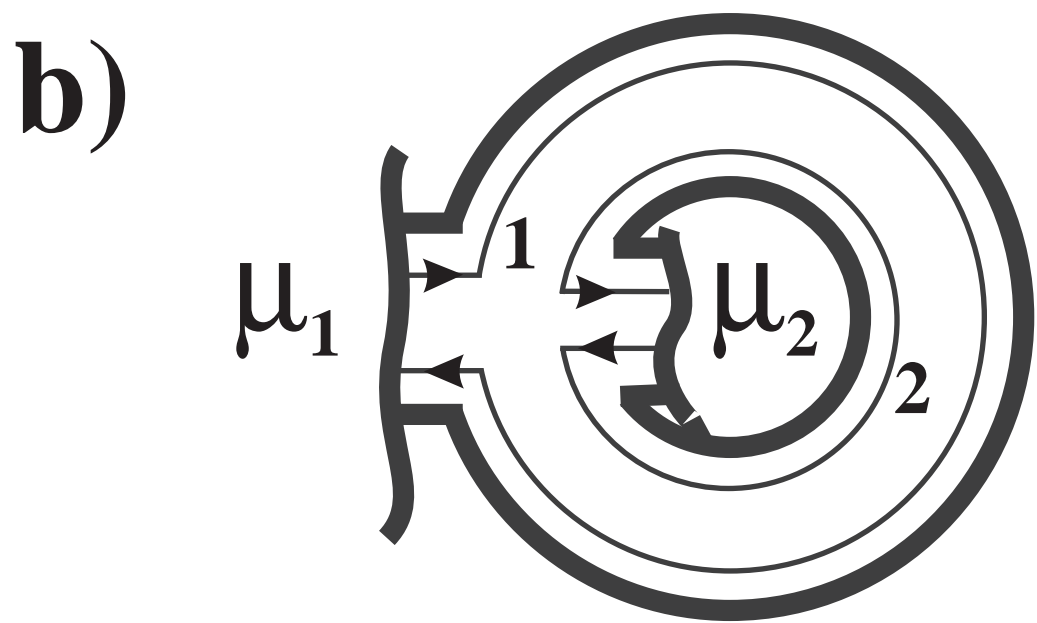
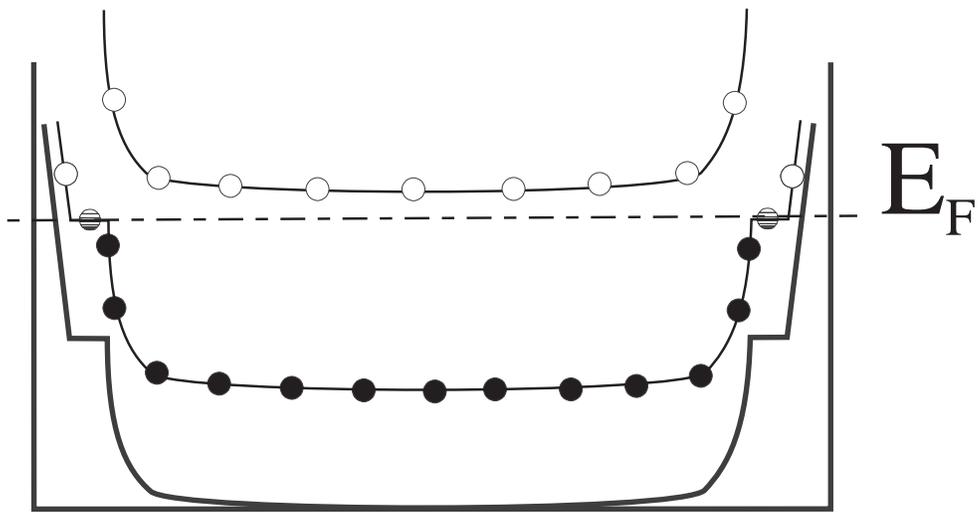
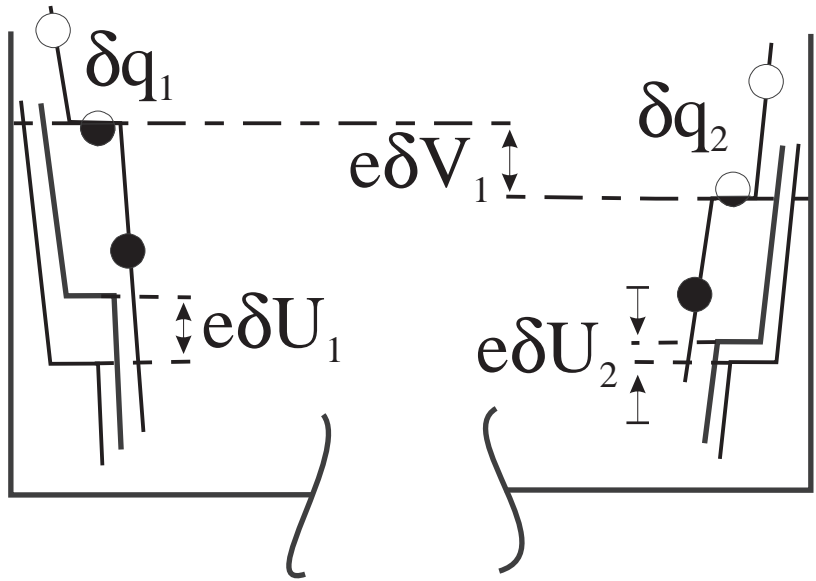
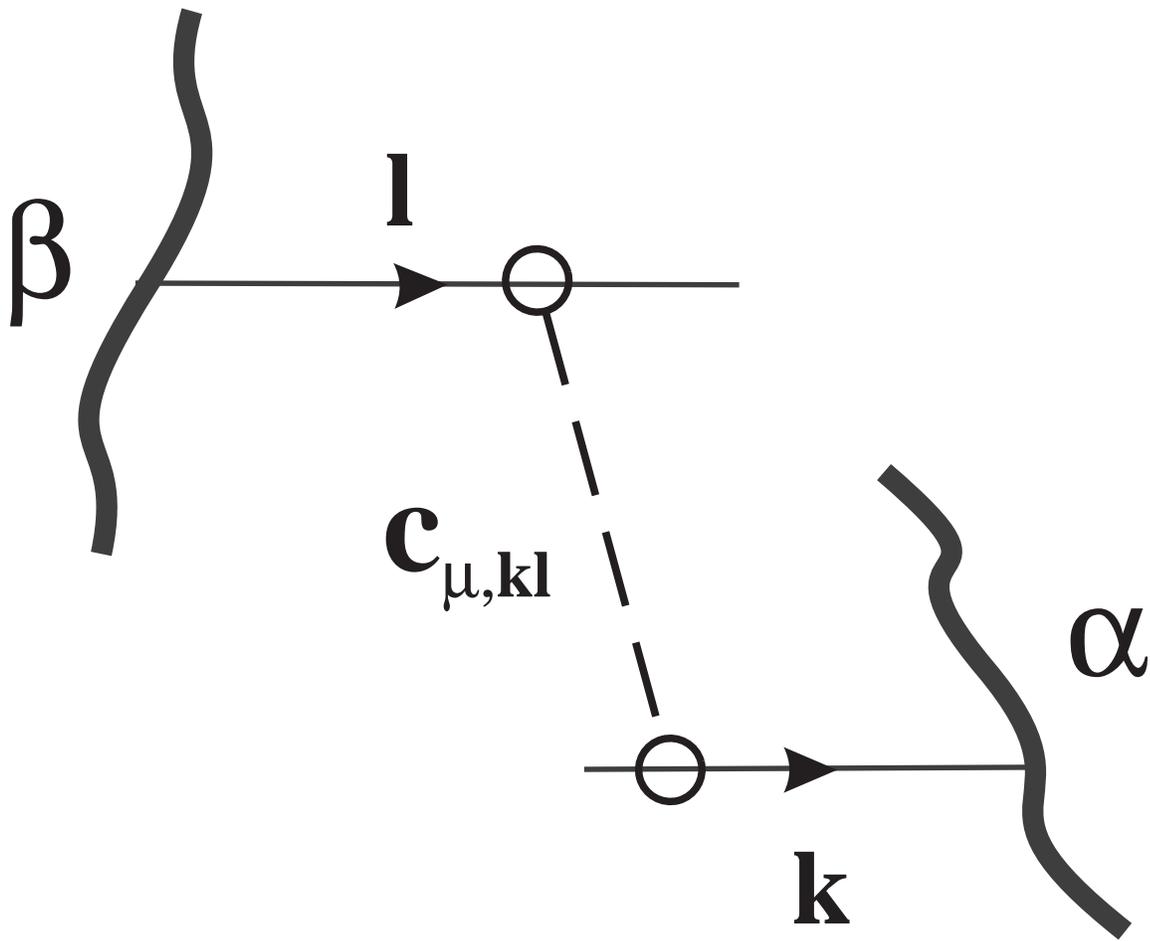
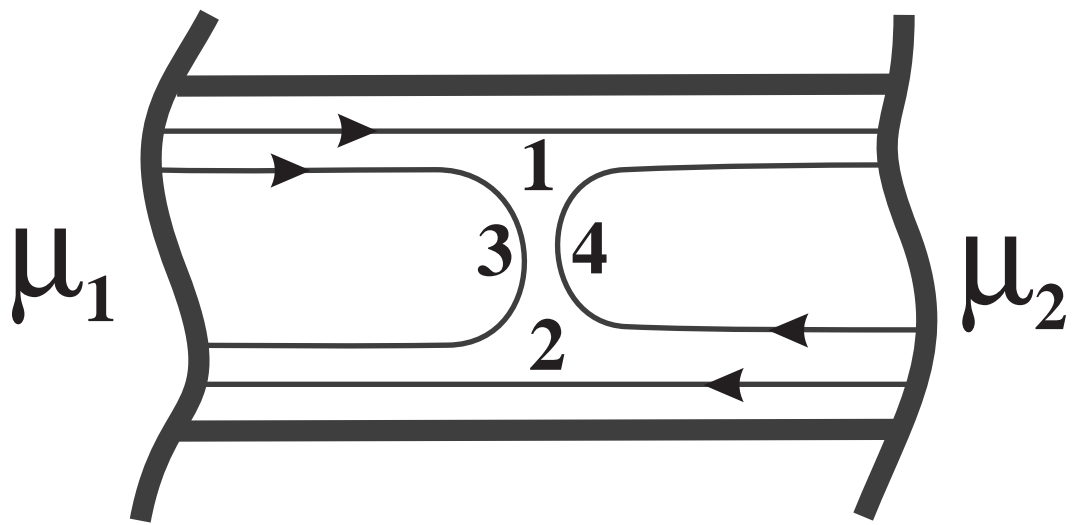
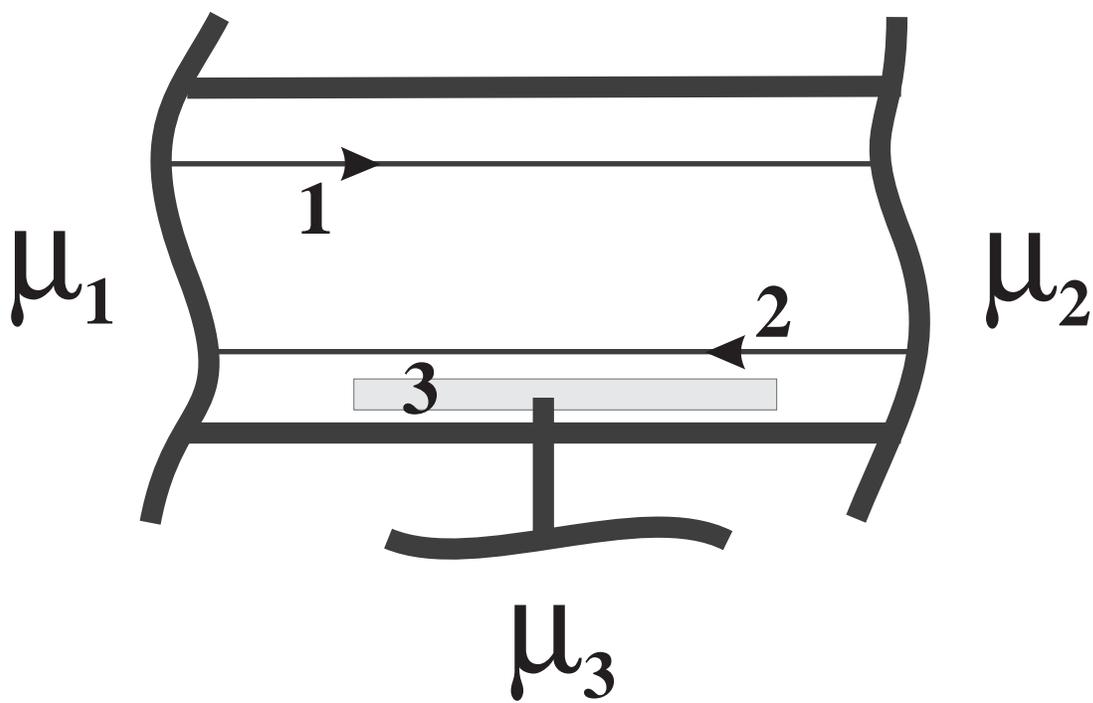
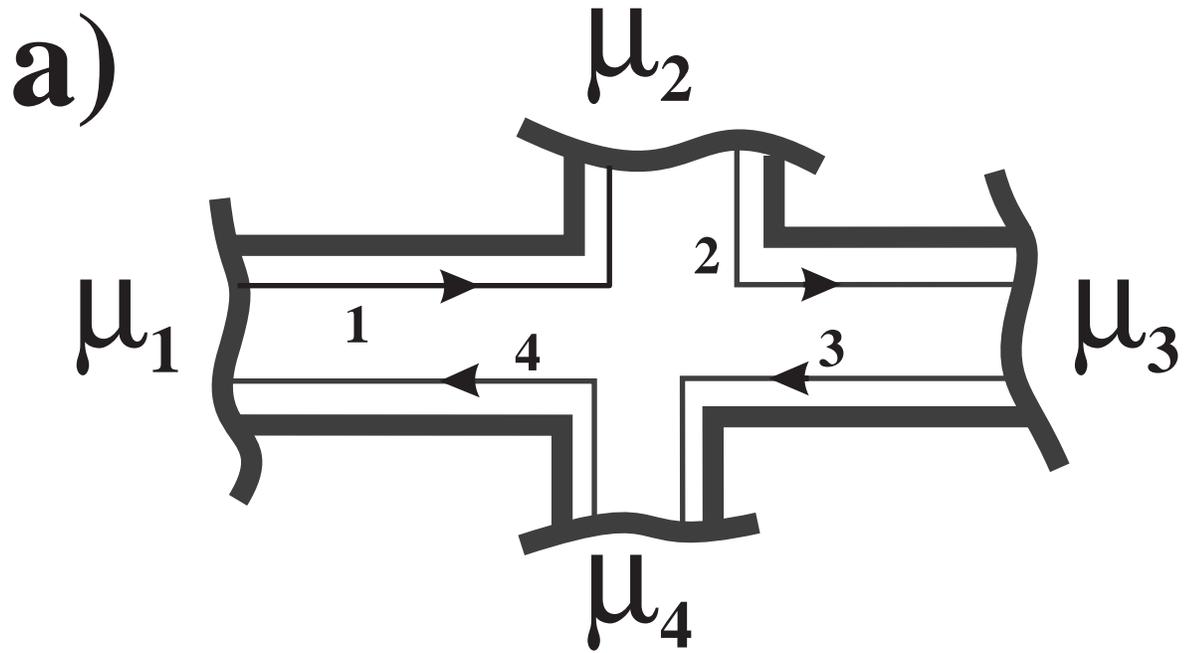
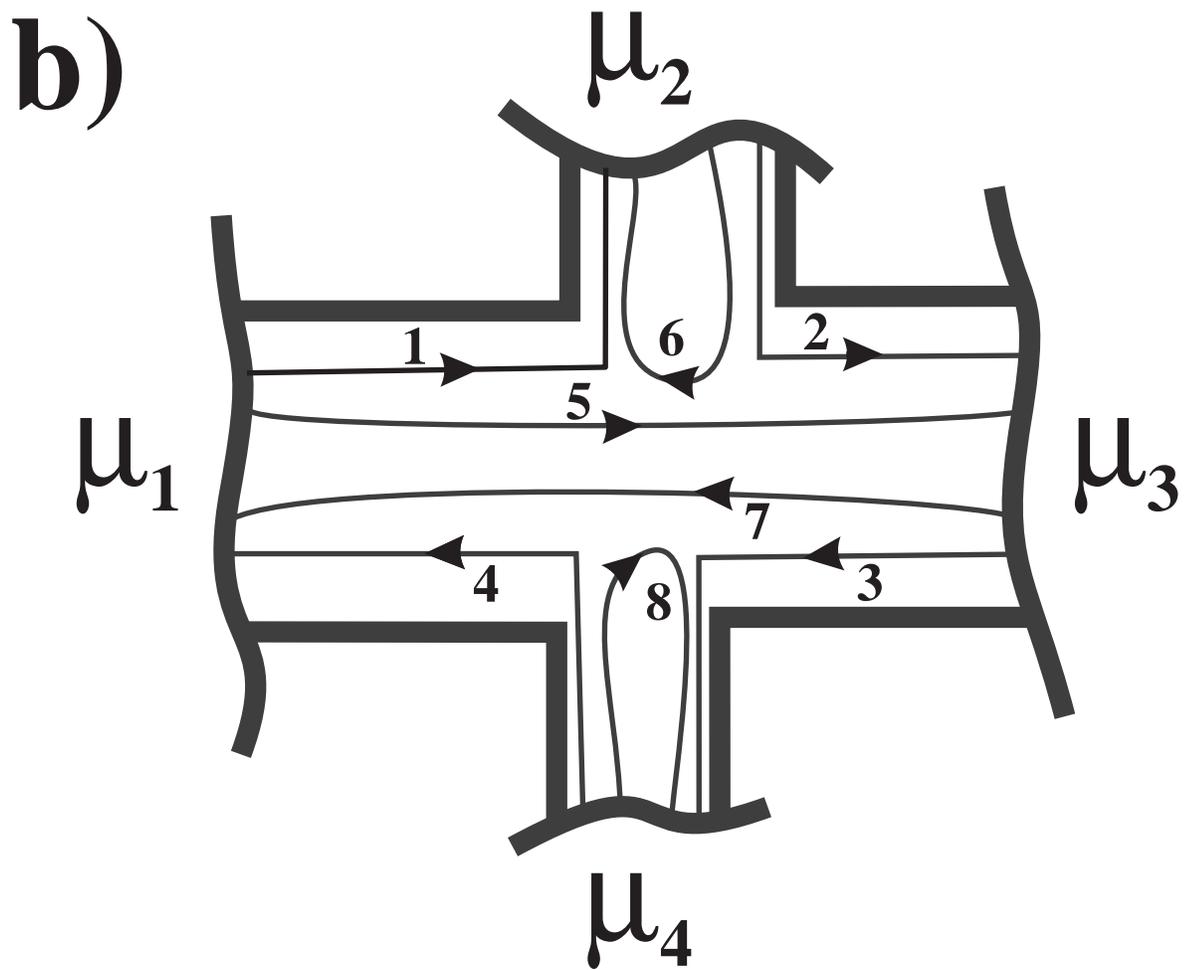
\end{document}